\documentclass[twocolumn, fleqn,10pt]{wlscirep}
\usepackage{multicol}
\usepackage{graphicx}
\usepackage{caption}
\usepackage{subcaption}
\usepackage[utf8]{inputenc}
\usepackage[T1]{fontenc}
\usepackage{breqn}
\usepackage{float}
\usepackage{comment}
\usepackage{amsmath}

\title{A Physics based Machine Learning Model to characterize Room Temperature Semiconductor Detectors in 3D}

\author[1,*]{Srutarshi Banerjee}
\author[2]{Miesher Rodrigues}
\author[1]{Manuel Ballester}
\author[2]{Alexander H. Vija}
\author[1]{Aggelos K. Katsaggelos}
\affil[1]{Northwestern University, 2145 Sheridan Road, Evanston, IL 60208, USA}
\affil[2]{Siemens Medical Solutions USA, Inc., Hoffman Estates, IL 60192, USA}

\affil[*]{srutarshibanerjee2022@u.northwestern.edu}

\begin{abstract}
Room temperature semiconductor radiation detectors (RTSD) for X-ray and $\gamma$-ray detection are vital tools for medical imaging, astrophysics and other applications. CdZnTe (CZT) has been the main RTSD for more than three decades with desired detection properties. In a typical pixelated configuration, CZT have electrodes on opposite ends. For advanced event reconstruction algorithms at sub-pixel level, detailed characterization of the RTSD is required in three dimensional (3D) space. However, 3D characterization of the material defects and charge transport properties in the sub-pixel regime is a labor intensive process with skilled manpower and novel experimental setups. Presently, state-of-art characterization is done over the bulk of the RTSD considering homogenous properties. In this paper, we propose a novel physics based machine learning (PBML) model to characterize the RTSD over a discretized sub-pixelated 3D volume which is assumed. Our novel approach is the first to characterize a full 3D charge transport model of the RTSD. In this work, we first discretize the RTSD between a pixelated electrodes spatially into $3$ dimensions - x, y, and z. The resulting discretizations are termed as voxels in 3D space. In each voxel, the different physics based charge transport properties such as drift, trapping, detrapping and recombination of charges are modeled as trainable model weights. The drift of the charges considers second order non-linear motion which is observed in practice with the RTSDs. Based on the electron-hole pair injections as input to the PBML model, and signals at the electrodes, free and trapped charges (electrons and holes) as outputs of the model, the PBML model determines the trainable weights by backpropagating the loss function. The trained weights of the model represents one-to-one relation to that of the actual physical charge transport properties in a voxelized detector.
\end{abstract}
\begin{document}

\flushbottom
\maketitle
%
%
\thispagestyle{empty}

\section*{Introduction}

RTSDs are used for X-ray and $\gamma$-ray detection over the last three decades in medical imaging, astrophysics, homeland security and other applications which require high performance detectors at low costs \cite{schlesinger2001cadmium, butler2008bio, del2009, johns2019room}. RTSDs do not have the requirement of cryogenic cooling systems and have high bandgap, high atomic number, high density which are desirable absorption of high energy photons \cite{knoll2010radiation}.  Additionally, high quality RTSDs with uniform and optimized charge transport properties are also desired. Typically, RTSDs with no polarization effect, excellent fabrication quality, high breakdown voltage, high drift velocity of charges and high energy resolution is desired. CZT used over several decades has the desired detection properties.

Pixelated CZT are used in pulse mode at high fluxes \cite{soldner2007dynamic, bale2008mechanism} and high performance has been shown at room temperatures. The yield and performance of high quality detector material is limited by presence of randomly distributed high concentrations of defects. CZT is also prone to polarization effects (at high photon fluxes $10^{6} mm^{-2} s^{-1}$) due to the build up of trapped charge in the crystal \cite{bale2008mechanism}. Recently developed high flux capable CZT \cite{iniewski2016czt} has been characterized in high-flux scenarios \cite{thomas2017characterisation}, and also been with intense light sources like LCLS XFEL \cite{veale2018cadmium} and ESRF synchrotron \cite{tsigaridas2019x}. Detrimental defects are also observed in CZT \cite{roy2019evaluation}, which are attributed to compositional inhomogeneity due to non-unity segregation coefficient of Zn \cite{zhang2011anomalous}, high concentration of secondary phases, Te inclusions and wall dislocations, and localized fluctuations of electric field \cite{veale2020characterization}. These defects acts as trapping centers and promotes non-uniformity in charge transport, thereby affecting the detector's performance adversely \cite{camarda2008polarization, bolotnikov2013characterization, veale2020characterization,  roy2021advances}. Thus, high degree of uniformity is required before further widespread use of CZT \cite{wilson201510, zambon2018spectral} in different applications. 

Characterization of CZT and other RTSDs have been attempted for the last several years. Thermal ionization energies of electron and hole traps were measured using thermoelectric emission spectroscopy (TEES) and thermally stimulated conductivity (TSC) measurements \cite{lee1999compensation}. Using microwave cavity perturbation techniques \cite{tepper2000contactless}, trap lifetime was determined in CdZnTe and HgI$_{2}$. In CdZnTe, the electron and hole traps were irradiated with $5$ MeV focused proton beam to generate electron-hole pairs and fill traps, which were later released by thermal re-emission. By excitation near the vicinity of the appropriate electrodes, electron and hole traps were distinguished \cite{medunic2005studying}. Analysis of simultaneous multiple peaks on TSC measurements showed deep trap levels in CdZnTe \cite{pavlovic2008identification}. TSC measurements further showed $9$ defect levels and irradiation-induced variations of these levels on CdZnTe:Al \cite{nan2012irradiation}. Using statistical model of charge collection efficiency based on known electron average trapping time, the average hole trapping time was derived \cite{rodrigues2011high}. By comparing the measured and simulated signals for holes as measured in the cathode, average hole de-trapping time was extracted \cite{blakney1967small, jung1999new, prokesch2010fast}. The change in carrier mobility due to effects of deep trap in CZT have been studied \cite{xu2014effects}. Indium doped CZT crystal with $13$ trap levels had been observed \cite{zaman2015characterization}. Using pulsed laser microwave cavity perturbation method selectively at the surface and in the bulk region of CZT, imperfections due to mechanical damage or adsorbed chemical species which lead to charge trapping or increase of leakage current have been characterized \cite{tepper2001investigation}. The influence of deposition techniques and type of metal contacts on the recombination and trapping defects at the metal-semiconductor interface has been studied \cite{zheng2012investigation}. The high flux capable CdZnTe and its uniformity has also been characterized \cite{veale2020characterization}. In literature, the defects and charge transport properties of electrons and holes are measured considering homogeneous behavior of the defects in the RTSD. However, the homogeneity of the material properties and repeatability of the defects and charge transport properties within a detector and across multiple detectors are unknown. Additionally, for high energy resolution and sub-millimeter position detection accuracy, a thorough in-depth sub-pixel level characterization of RTSDs in required. However, this approach requires numerous novel sophisticated experiments, skilled manpower and is traditionally hugely time consuming. On the other hand, deployment of such detector arrays requires precise characterization of individual detectors, and knowledge of the defects in a 3D manner.

In this work, a PBML approach is developed for 3D sub-pixel characterization of the RTSD. Machine learning has been tremendously popular in the last decade with several novel works virtually across every major science and engineering discipline. In the recent past Deep Learning (DL) \cite{alom2018history} gained popularity, in particular the classes of Convolutional Neural Networks (CNNs) \cite{krizhevsky2012imagenet}, \cite{karen2014deep}, Recurrent Neural Networks (RNN) \cite{gers2000recurrent}, \cite{gers2002learning} and Generative Adversarial Networks (GAN) \cite{goodfellow2014generative} being the most popular architectures. Recent focus has been on applying machine learning to physics based systems, material science, drug discovery and others. Integrating physics-based modeling and Machine Learning is becoming more popular over the years \cite{willard2020integrating}, \cite{karniadakis2021physics}. Solving problems in physics governed by PDEs using Neural Networks has been done \cite{khoo2019solving, han2018solving}. DeepONets \cite{lu2021learning} have been demonstrated as a powerful tool to learn nonlinear operators in a supervised data-driven manner. Two-dimensional wave equation is modeled as a Recurrent Neural Network \cite{manuel_RNN}. $2$D Poisson Equation has been solved with a Physics Informed Neural Networks \cite{markidis2021old}. In most of these PBML approaches, relatively simpler PDEs have been solved. However, the charge transport in a RTSD has multitude of coupled PDEs \cite{miesher_phd} involving charge drift, trapping, detrapping and recombination of electrons and holes.

In this paper, we develop the PBML approach for characterizing the RTSD in 3D, from the 1D physical models as described in our previous works \cite{Siemens_pap1, Siemens_pap2, Srutarshi_PhD, banerjee2023machine, Siemens_IP1, Siemens_IP2, Siemens_IP3}. \textbf{We propose to reap the benefits of machine learning in characterizing radiation detectors in 3D at a sub-pixel resolution, which to the best of our knowledge is the first contribution in this area.} 

Our novel 3D PBML model is derived from the physical charge transport equations for both electrons and holes. Using the PBML approach, we aim to solve for the electron and hole drift coefficients as well as the material defects such as trapping, detrapping and recombination lifetimes for electrons and holes. Multiple trapping centers are also considered in this approach. Compared to classical methods, our PBML model identifies the defects at a high spatial resolution of $100$ $\mu$m. The input to the model consists of electron-hole pairs at different positions and the output from the model consists of signals at the electrodes along with the free and trapped electron and hole charges in the voxels over time. The 3D RTSD volume is spatially discretized into voxels by virtually segmenting along x-axis, y-axis and z-axis. The physical charge transport equations are considered in each voxel. Our PBML model does not only consider the charge transport normal to the electrode surfaces in a pixel but also the second order effects due to charge motion in the lateral direction. Compared to a traditional Convolutional Neural Network, Recurrent Neural Network or Fully Connected Networks, which typically contains millions of trainable weights, our model is designed to have the same number of trainable weights as the number of unknowns in the physical equation of a discretized RTSD. Conceptually, prior knowledge as described by the physical laws and the physical process is used in this model. Moreover, the trained model weights are related one-to-one with the detector material properties as dictated by the discretized physical charge transport equations. 

This proposed novel PBML model for 3D RTSDs in sub-pixel domain aims to solve the following problems currently plaguing the characterization of radiation detectors with a reasonable detection area for wide scale implementation in medical imaging and security applications:
\begin{itemize}
    \item Fine characterization of detector material properties spatially in 3D in a fast and efficient way.
    \item Micron level defect identification and characterization in sub-pixelated volume.
    \item Application of corrections to these RTSDs at sub-pixel level.
\end{itemize}

\section*{Results}

\begin{figure}
  \centering
  \includegraphics[width=0.4\linewidth]{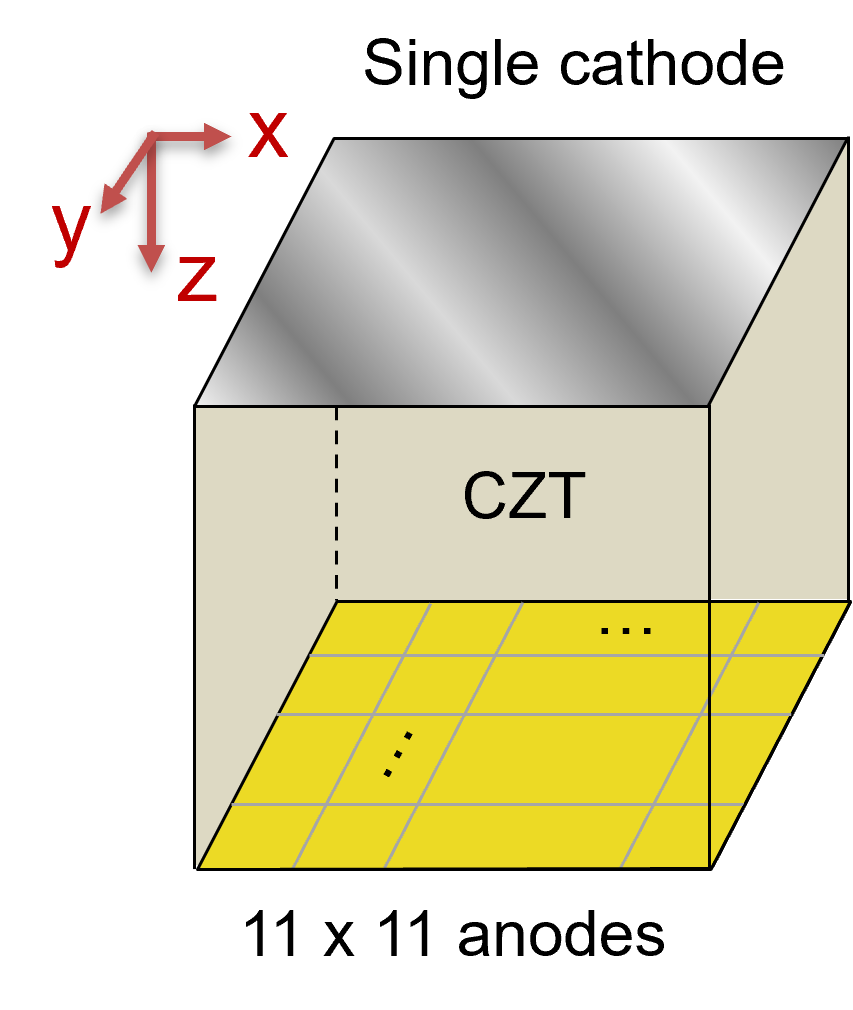}
  \caption{Pixelated CdZnTe detector. Anode is pixelated and a single cathode.}
\label{chap_3D_electrode_config}
\end{figure}

In solid-state detectors, electrons and hole transport properties play a significant role in selecting detectors for any applications. When high energy X-rays or $\gamma$ photons hit the CdZnTe detector, it undergoes Photoelectric effect, Comption Scattering or Pair Production and deposit energy in the detector creating electron-hole pairs in these processes. The Shockley-Ramo theorem states that the moving charge within a detector induces charge on the electrodes of the detector which are extracted \cite{he2001review}. The weighting potential when a point charge is at a particular location determines the charge (or current) induced in the electrodes. In our study, we consider a single large electrode (cathode) and a pixelated anode ($11 \times 11$) as shown in Fig. \ref{chap_3D_electrode_config}.

The weighting potential can be calculated numerically using Finite Element Method (FEM) with ANSYS Maxwell software by solving the Poisson equation. The weighting potential of the $121$ anode pixels are not identical, and there are some variation in the weighting potential of central pixels, edge pixels and corner pixels. Moreover, the weighting potential under one pixel also shows slight lateral variation \cite{chen2020intrinsic}.  

\begin{figure}
  \centering
  \includegraphics[width=0.7\linewidth]{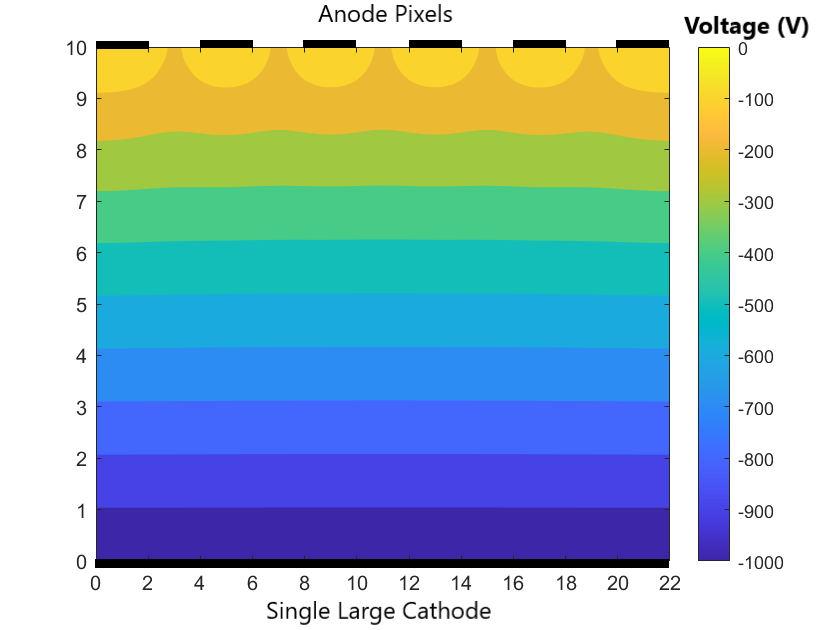}
  \caption{Variation of electric field between Cathode and Anode Pixels with a potential difference of $1000$ V for GaAs crystal.}
\label{chap_3D_det_full_m}
\end{figure}

\begin{figure}
  \centering
  \includegraphics[width=0.7\linewidth]{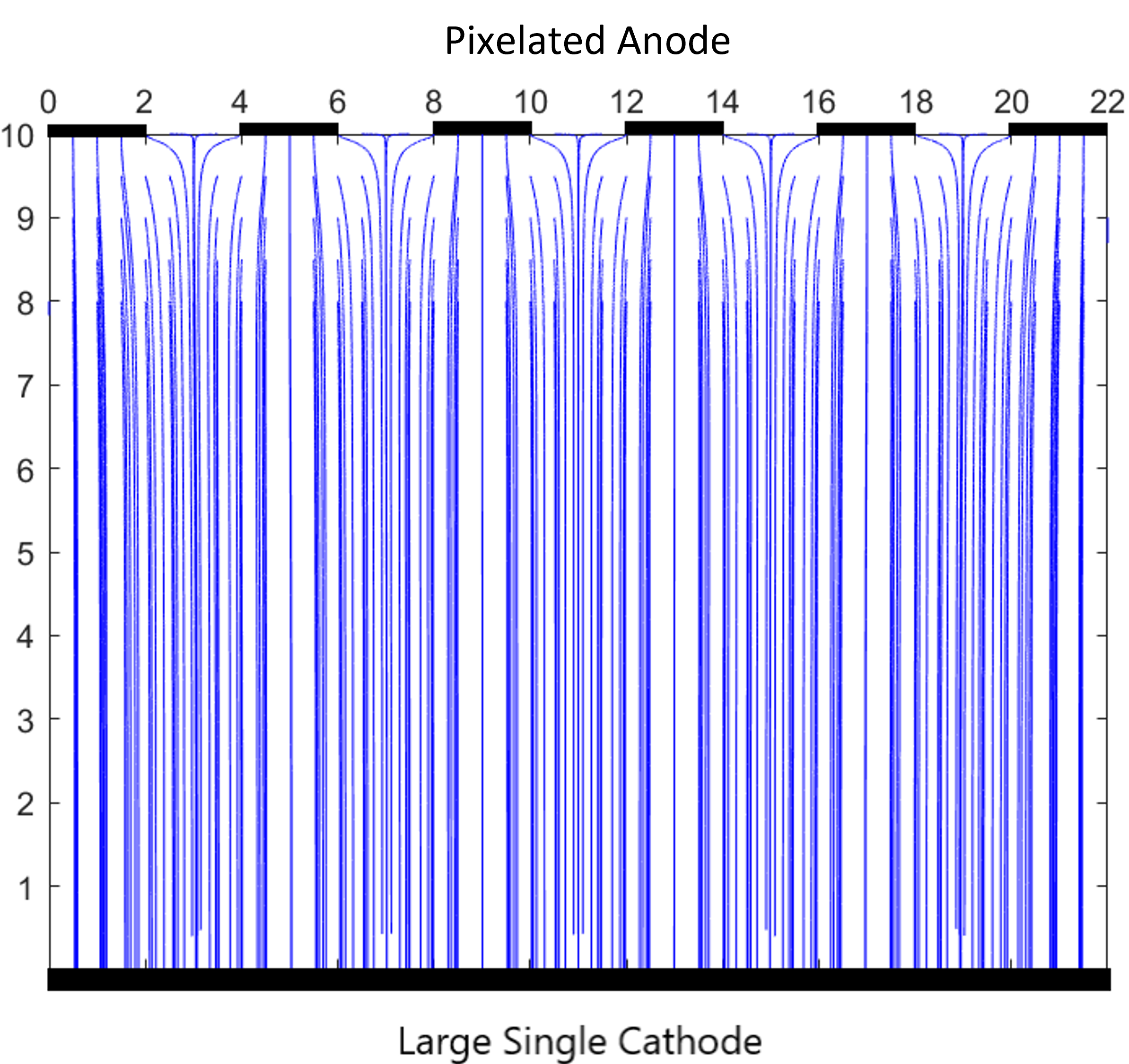}
  \caption{Variation of electric field between the Anode Pixels and Cathode with GaAs crystal. The electric field becomes non-uniform near the Anode Pixels.}
\label{chap_3D_e_field_near_anode}
\end{figure}

Additionally, the  variation of voltage along the volume of the detector in 2D as shown in Fig. \ref{chap_3D_det_full_m} for a case of $6$ pixelated anodes and a large cathode on the opposite end at a gap of $10$ mm. The length of the cathode is $22$ mm. The simulation for the electric field and voltage are done for GaAs RTSD which has a relative permeability $\mu_{r} = 12.9$ using COMSOL Multiphysics software. The voltage is uniform in most of the region between anodes and the cathode. For those electrons and holes under one pixel, the charges will drift perpendicularly to their corresponding electrodes. However, the charges in between the anode pixels will have different drift behavior \cite{chen2020intrinsic}. The non-uniformity of the voltage is shown in Fig. \ref{chap_3D_det_full_m} near the anodes. The corresponding electric field is non-uniform which is shown in Fig. \ref{chap_3D_e_field_near_anode}. In most of the region, the electric field is perpendicular to the cathode and anode pixels. However, near the anode, the electric field bends towards the nearest anode. The non-uniform electric field in between the anode pixels is shown in Fig. \ref{chap_3D_e_field_near_anode}. In the region of uniform electric field, the motion of electrons and holes will drift perpendicular to the electrodes. However, the non-uniform electric field under the gap will bend the trajectory of electrons closer to the nearest anode. The electrons in the gap between the anode pixels will be collected to the closest anode pixel. Thus, the behavior of the electrons will change close to the pixelated anodes.

Additional effects of charge trapping, pixel jumping, non-uniform electric field due to different charge concentrations may also be  present in the detector. The trapping, detrapping and recombination is governed by Schockley-Read-Hall Theory \cite{shockley1952statistics}, \cite{hall1952electron}.The trapping and detrapping lifetimes dictate whether the defects induce short or long term trapping of charges in the detector. Signals collected at the anodes and cathode arise out of the movement of charges. We consider the CdZnTe detector with $2$ trapping centers for holes and $1$ trapping center for electrons.

The data for training the proposed 3D PBML model has been generated in MATLAB using the charge transport equations \cite{Siemens_pap1, miesher_phd}, by defining the drift coefficients, trapping, detrapping and lifetimes for electrons and holes as pre-defined parameters $\mu_{e}$, $\mu_{h}$, $\tau_{eT}$, $\tau_{eD}$, $\tau_{hT1}$, $\tau_{hD1}$, $\tau_{hT2}$, $\tau_{hD2}$, $\tau_{e}$, and $\tau_{h}$ respectively, alongwith electric field in the detector. The classical model has been created by spatially discretizing the detector. For charge input at different voxels in the classical model, the signals are generated at the electrodes. Additionally, the free and trapped electron and hole charges are computed for each time step, with the total time steps defined \emph{a priori}. The input-output data for training the learning-based model consists of the input electron-hole pair injected in the detector at a known voxel position alongwith the signals, free and trapped electron and hole charges in different spatial voxels of the classical model over different time steps in the simulation.

\subsection*{3D Learning Model}
\label{3D_Chap_Learning_Model}
\begin{figure}
  \centering
  \includegraphics[width=\linewidth]{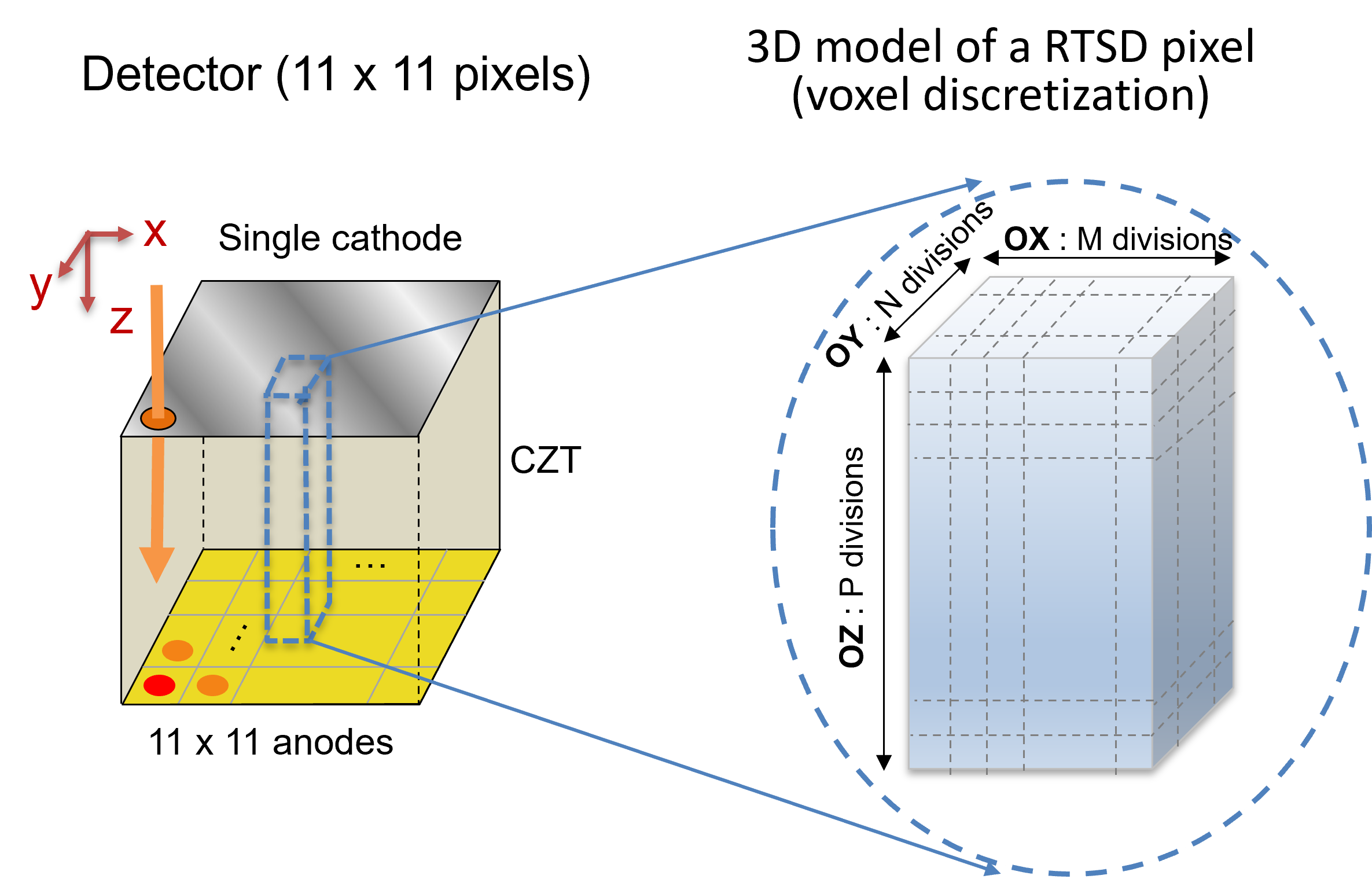}
  \caption{CZT detector with 121 pixelated anodes at one end and a single planar cathode on the opposite end (left). A single pixel is further discretized into subpixels (right).}
\label{121_pixels_single_pixel}
\end{figure}

The configuration of CdZnTe detector in our PBML model consists of a CdZnTe crystal with a single planar cathode on one end and several anode pixels ($11 \times 11$) on the other end as shown in Fig. \ref{121_pixels_single_pixel} (left). Out of the $121$ anode pixels, we just use the cuboidal volume of the crystal under the central anode pixel and the cathode on the other end, and subdivide this volume into voxels as shown in Fig. \ref{121_pixels_single_pixel}. In the x-direction (\textbf{OX}), we divide into $M$ divisions, in the y-direction (\textbf{OY}), we subdivide into $N$ divisions and in z-direction (\textbf{OZ}), we subdivide into $P$ divisions. Thus, the total voxels are $M\times N \times P$. Since the single anode pixel is subdivided in x and y-direction into $M$ and $N$ divisions, we consider the number of subpixels as $M \times N$.

The material properties of the CdZnTe for electrons and holes, such as $\mu_{e}$, $\mu_{h}$, $\tau_{eT}$, $\tau_{eD}$, $\tau_{hT1}$, $\tau_{hD1}$, $\tau_{hT2}$, $\tau_{hD2}$, $\tau_{e}$, and $\tau_{h}$ are considered at each voxel $d$ as parameters independent of the neighboring voxels. These refer to the drift coefficients for electrons and holes, electron trapping lifetime, electron detrapping lifetime, hole trapping 1 lifetime, hole detrapping 1 lifetime, hole trapping 2 lifetime, hole detrapping 2 lifetime, electron lifetime and hole lifetime respectively. Thus, for a voxel $d$, the material properties are expressed as  $\mu_{e,d}$, $\mu_{h,d}$, $\tau_{eT,d}$, $\tau_{eD,d}$, $\tau_{hT1,d}$, $\tau_{hD1,d}$, $\tau_{hT2,d}$, $\tau_{hD2,d}$, $\tau_{e,d}$, and $\tau_{h,d}$. We refer to these discretized properties as trainable weights spatially distributed in the CdZnTe detector. Each of the material properties (such as $\tau_{eT}$) are thus converted into $M \times N \times P$ discrete values. For each of these properties or coefficients (referred here as $\tau$ in general), we compute the number of charge particles (electrons or holes) remaining at that particular level as $N_{left} = N_{0}e^{-t_{1}/\tau}$, where $N_0$ and $N_{left}$ are the number of charged particles at a particular energy level at time $t=0$ and $t=t_{1}$ respectively. In our learning model with $100$ voxels, we used the $t_{1} = 10$ns time steps. For a given $\tau$, we can compute the fraction of charges transitioning from one energy level to another, which is the probability of transition of charges. We consider all these probabilities as trainable weights in the PBML model. For example in voxel $d$, $w_{eD,d}$ is the probability of the electron getting detrapped from the trapped level to the conduction band.

\begin{figure}
  \centering
  \includegraphics[width=\linewidth]{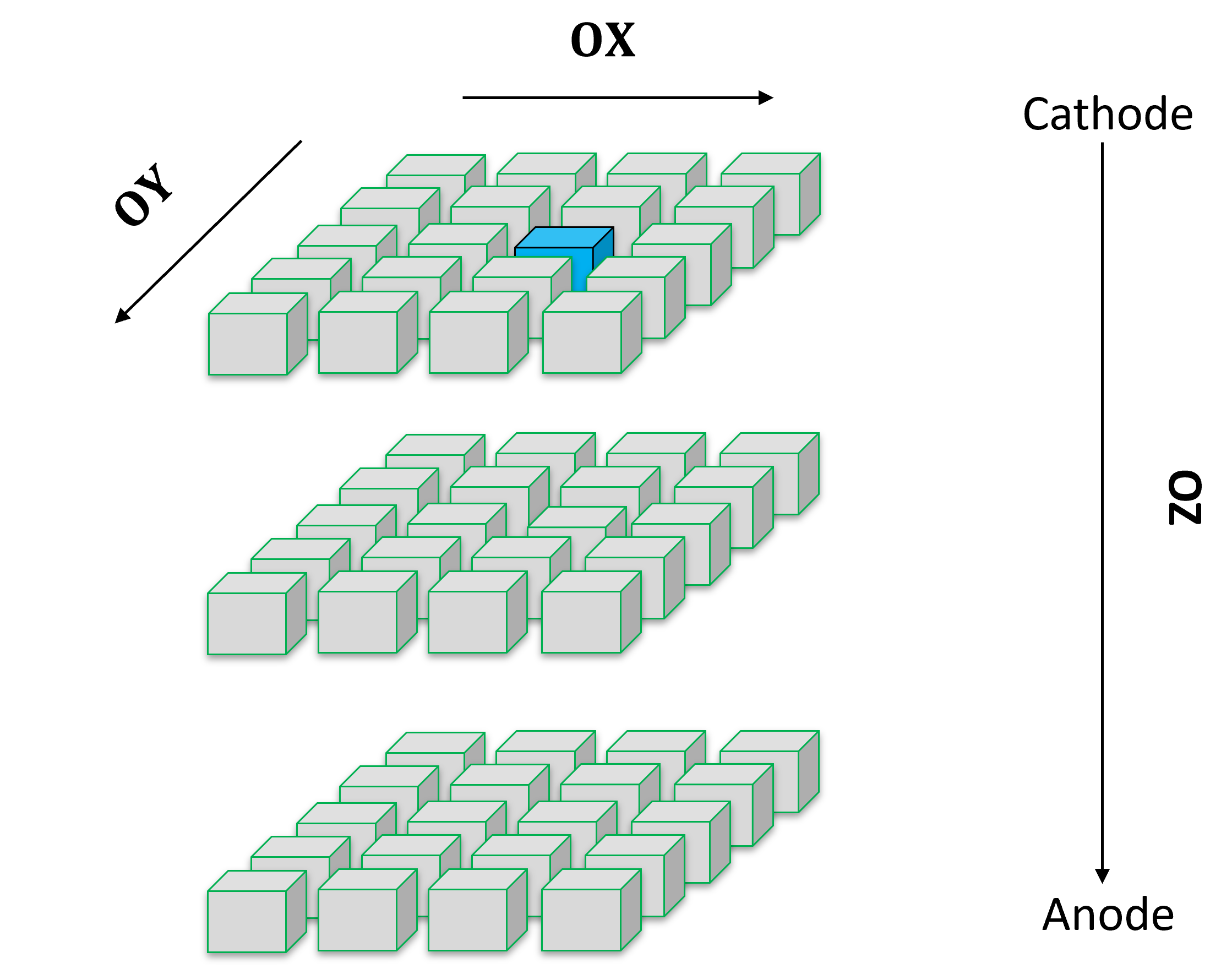}
  \caption{Electrons at a particular voxel position at time $t = 0$.}
\label{ch_e_t_0}
\end{figure}

\begin{figure}
  \centering
  \includegraphics[width=\linewidth]{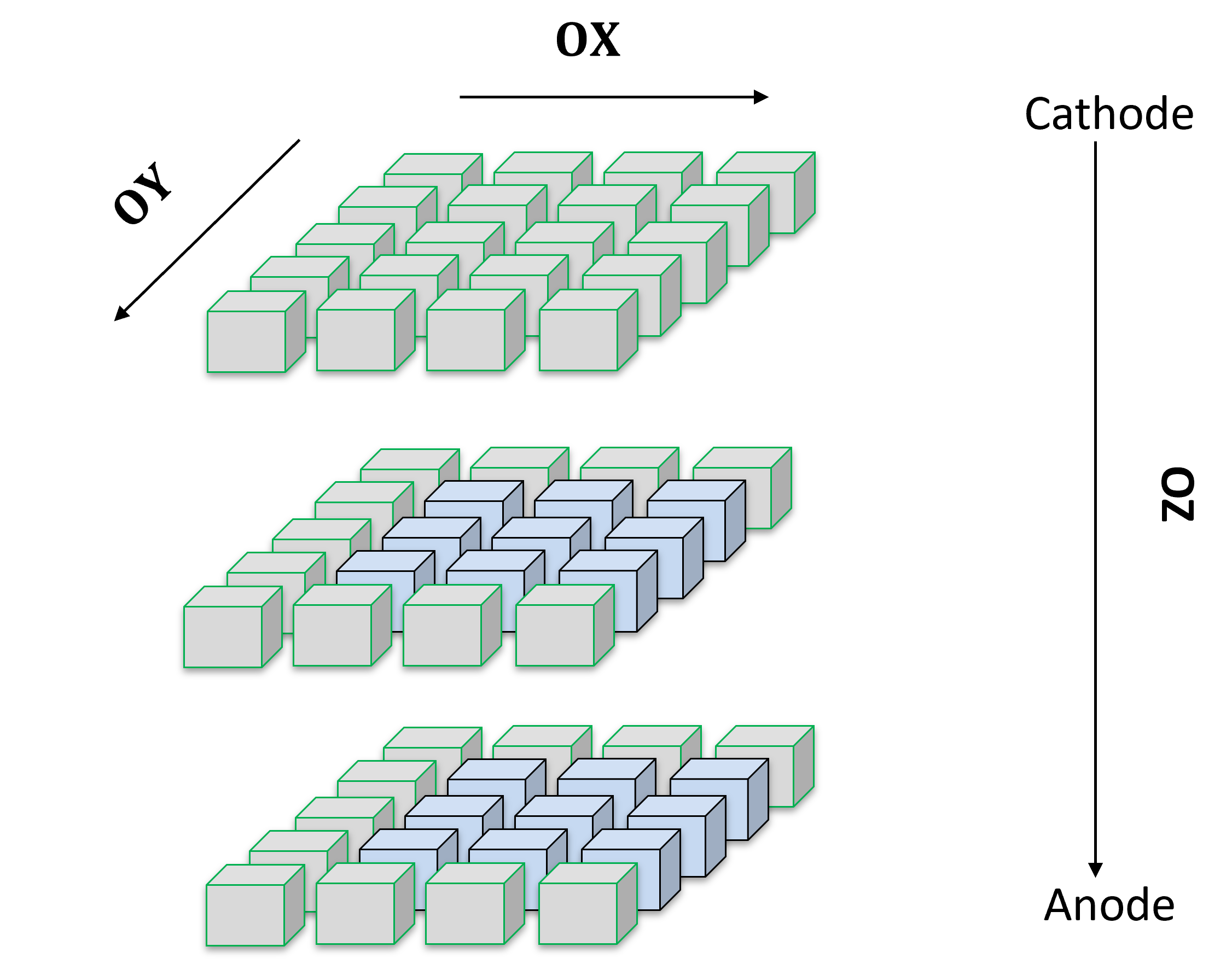}
  \caption{Electron drift in the neighboring voxels at time $t = t_{1}$.}
\label{ch_e_t_t1}
\end{figure}

The motion of charges is taken into account by considering non-uniform electric field between the electrodes (cathode and anode) as shown in Figs. \ref{chap_3D_det_full_m} and \ref{chap_3D_e_field_near_anode}. The charge transfer thus occurs not only in z-direction ($\textbf{OZ}$) which is perpendicular to the electrodes of opposite polarity but also in x and y-direction ($\textbf{OX}$) and ($\textbf{OY}$) as well. Fig. \ref{ch_e_t_0} shows  electrons at a particular voxel at time $t=0$. Due to the non-uniform electric field, the electrons move in 3D direction. However, the motion in z-direction ($\textbf{OZ}$) is much significant compared to the motion in the ($\textbf{OX}$) and ($\textbf{OY}$) directions since the electric field is much stronger in the z-direction. When electron moves in z-direction, we assume the motion of the electrons in the x-y plane to be constrained in the immediate neighboring locations. For a voxel position of $(i,j,k)$, the electron drifts to immediate neighboring locations of $(i+1,j)$, $(i-1,j)$, $(i,j+1)$, $(i,j-1)$, $(i+1,j+1)$, $(i-1,j-1)$, $(i+1,j-1)$, $(i-1,j+1)$ in the x-y plane. Fig. \ref{ch_e_t_t1} shows the electron drift in 3D during a time step. From the z position of $k$, the electrons move in the immediate neighboring voxels having z position $k+1$ and $k+2$. The holes move in the opposite direction from Anode to Cathode due to its opposite polarity. 

Once the charge, for example holes drift from neighboring voxels to a voxel $(i,j,k)$ at time $t$, it is added to the existing holes in that voxel from the previous time $t = t-1$ to form net hole charges. Equation \ref{eqn:eqlabel8} shows the phenomena of the drift of holes during the time $t$, forming total hole charge $q^{t,(i,j,k)}_{h,int}$. A fraction of this net hole charge $q^{t,(i,j,k)}_{h,int}$ then combines with the intrinsic electron concentration in bulk of the material at voxel $(i,j,k)$ with weights $w_{hRec,(i, j, k)}$ as shown in Eqn. \ref{eqn:eqlabel9}. The resulting free holes after recombination is $q^{t,(i,j,k)}_{h,int1}$ as shown in Eqn. \ref{eqn:eqlabel9a}. The concentration of holes in the $r$ th trap center increases based on the fraction $w_{hT,r,(i,j,k)}$ of free holes $q^{t,(i,j,k)}_{h,int1}$ getting trapped and the fraction $w_{hD,r,(i,j,k)}$ of holes in that trap center $\tilde{q}^{ t,(i,j,k)}_{h,r}$ which is detrapped. We consider $R$ trapping centers for holes. $q^{t,(i,j,k)}_{h,mob}$ in Eqn. \ref{eqn:eqlabel10} is the net free / mobile holes considering the phenomena of trapping and detrapping in the $R$ different hole trapping centers of the RTSD. A fraction of this free holes $q^{t,(i,j,k)}_{h,mob}$ drifts with weight $w_{h,(i,j,k)}$ to the neighboring voxels as shown in Eqn. \ref{eqn:eqlabel11}. Similar equations are also valid for electrons which now moves from Anode to Cathode due to opposite polarity.

\begin{figure}
  \centering
  \includegraphics[width=\linewidth]{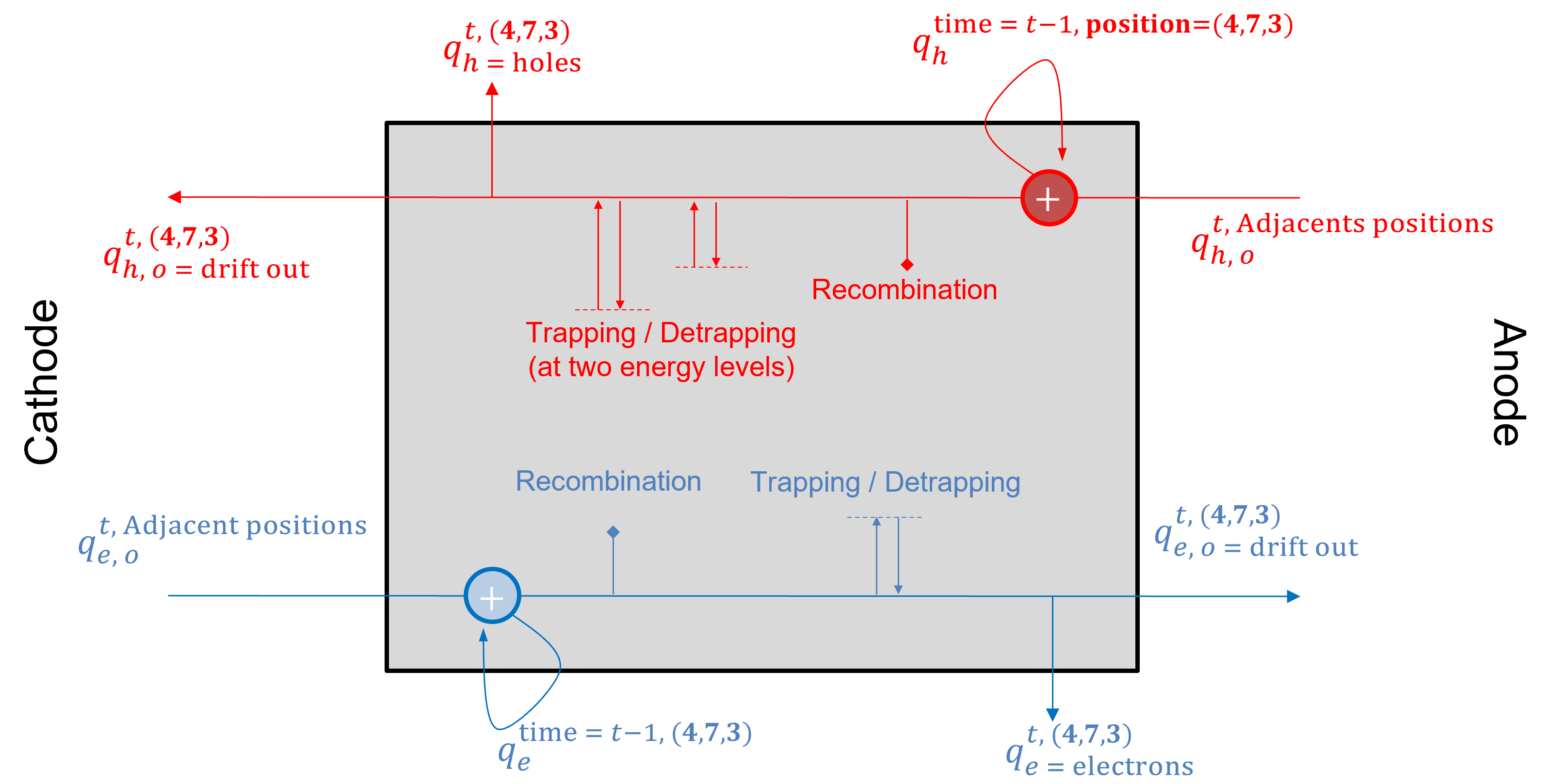}
  \caption{Operation in Voxel $(4,7,3)$ at time $t$.}
\label{Vox_op_3D}
\end{figure}

For illustrative purposes, Fig. \ref{Vox_op_3D} shows the operation in a voxel $(4,7,3)$ at time $t$. At any time $t-1$, the charge in that voxel is $q^{t-1,(4,7,3)}_h$ for holes and $q^{t-1,(4,7,3)}_e$ for electrons. The electrons from adjacent voxels in 3D drift at time $t$ and are added onto the existing charge. Some of the net electrons gets recombined in the bulk of the crystal with the intrinsic hole concentration in the bulk of the material based on the electron lifetime probability $w_{e,(4,7,3)}$. The electrons then get trapped with probability $w_{eT,(4,7,3)}$ to the trap level in the detector and detraps electrons back as excess electron concentration over bulk with a probability of $w_{eD,(4,7,3)}$. Following the 3D electric field, a fraction of the electron charges may be left behind in the voxel (4,7,3) while the remaining fraction drifts to the neighboring voxels. Exactly same operations are repeated for holes, with the holes drifting from anode to cathode. 

\begin{dmath}
\label{eqn:eqlabel8}
q^{t,(i,j,k)}_{h,int} = q^{t-1,(i,j,k)}_{h} + (q^{t,(i+1,j,k)}_{h,o} + q^{t,(i,j+1,k)}_{h,o} + q^{t,(i,j,k+1)}_{h,o} + q^{t,(i+1,j,k+1)}_{h,o} + q^{t,(i,j+1,k+1)}_{h,o} + \cdots) 
\end{dmath}

\begin{dmath}
\label{eqn:eqlabel9}
q^{t,(i, j, k)}_{h,Rec} = w_{hRec,(i, j, k)} \times q^{t,(i, j, k)}_{h,int}
\end{dmath}

\begin{dmath}
\label{eqn:eqlabel9a}
q^{t,(i, j, k)}_{h,int1} = q^{t,(i, j, k)}_{h,int} - q^{t,(i, j, k)}_{h,Rec}
\end{dmath}

\begin{equation}
\label{eqn:eqlabel10a}
\begin{split}
\tilde{q}^{ t,(i,j,k)}_{h,r}= \tilde{q}^{ t,(i,j,k)}_{h,r} + (w_{hT,r,(i,j,k)} \times {q}^{ t,(i,j,k)}_{h,int1}) \\
- (w_{hD,r,(i,j,k)} \times \tilde{q}^{ t,(i,j,k)}_{h,r})
\end{split}
\end{equation}

\begin{equation}
\label{eqn:eqlabel10}
\begin{split}
q^{t,(i,j,k)}_{h,mob} = q^{t,(i,j,k)}_{h,int1} \times (1 - w_{hRec,(i,j,k)} - \Sigma_{r=1}^{R} w_{hT,r,(i,j,k)}) \\ + \Sigma_{r=1}^{R}(\tilde{q}^{ t,(i,j,k)}_{h,r}\times w_{hD,r,(i,j,k)})
\end{split}
\end{equation}

\begin{dmath}
\label{eqn:eqlabel11}
q^{t,(i,j,k)}_{h,o} = w_{h,(i,j,k)} \times q^{t,(i,j,k)}_{h,mob}
\end{dmath}

\begin{figure}
  \centering
  \includegraphics[width=\linewidth]{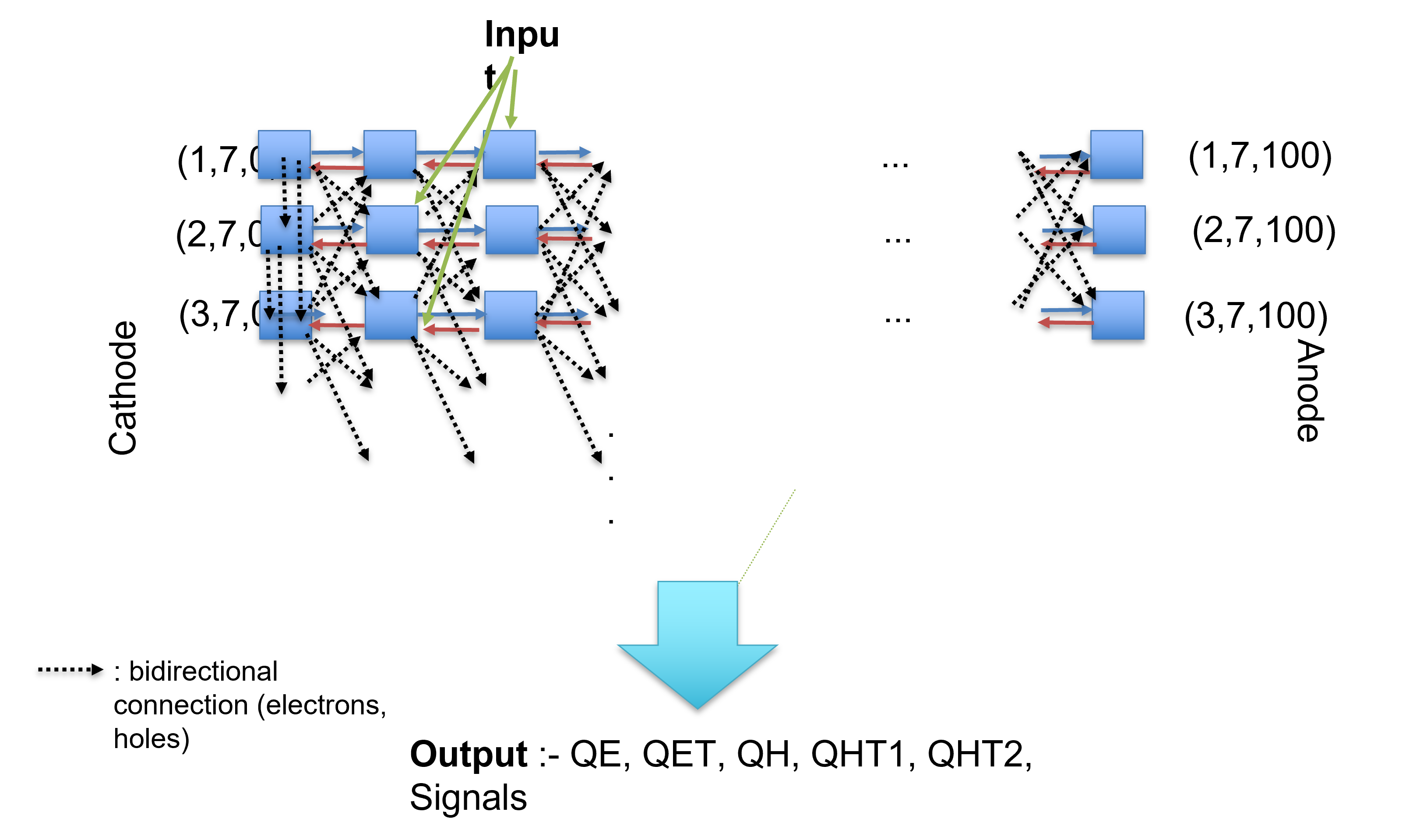}
  \caption{Interconnections among voxels in the 3D model. Each blue squares represents a Voxel shown in Fig. \ref{Vox_op_3D}.}
\label{Model_interconnects}
\end{figure}

The mathematical operations in each voxel in this 3D PBML model is also described in simplified voxelized 1D models \cite{Siemens_pap1, Siemens_pap2}. In our PBML model of charge transport the voxels are connected in a bidirectional manner in 3D to form a complicated PBML model. A simplified representation of the model is shown in Fig. \ref{Model_interconnects}. The left of the model has a planar cathode and the right consists of sub-pixelated anodes. The RTSD between the cathode and a anode pixel is subdivided into 3D voxels as shown with blue squares in Fig. \ref{Model_interconnects}. The input electron-hole charge pair injections can be in any of these voxels which depends on the position of interaction of high energy photons with the RTSD. The electrons drift towards the anode through the neighboring voxels in the z-direction (of Fig. \ref{ch_e_t_0} and Fig. \ref{ch_e_t_t1}) as shown by blue lines in the figure. On the other hand, the red lines show the drift of holes from one voxel to the neighboring voxels towards the cathode in the negative z-direction (of Fig. \ref{ch_e_t_0} and Fig. \ref{ch_e_t_t1}). The bidirectional connection shown by dotted lines indicate the movement of electrons and holes in the lateral direction (\textbf{OX} and \textbf{OY}) which basically represents lateral ($2$ nd order) drift of charges in x-y plane. 

The model is trained with input-output pairs of data. The input data is the positions of injected electron-hole pair and the output data are the signals obtained at the electrodes along with the electrons and holes (free and trapped) in each of the voxels over time. The weighting potential as defined by the electrode and detector configuration is non-uniform over the sub-pixels. During training the model, the loss function is computed as the sum of the squared errors between the signals at the electrodes and charges in the voxels compared to the ground truth signals along with the $error^{2}_{voltage}$. The overall loss function for training this PBML model is shown in Eqn. \ref{eqn:eqlabel_LF_3D}. The loss function is shown for $Z$ trapping centers for electrons and $R$ trapping centers for holes for a general model with several trapping centers for electrons and holes. However, we perform simulation experiments considering CZT detector with $2$ trapping centers for holes and $1$ trapping center for electrons \cite{lee1999compensation}, \cite{miesher_phd}. In the loss function, the errors due to the signals and voltage are grouped together, free and trapped electron charges are grouped together, and, free and trapped charges due to holes are grouped together with weighting terms $k$, $l$ and $n$ respectively. Clearly, as $k$, $l$ and/or $n$ are varied, the errors due to those terms vary. The higher the value of these parameters, the lower the errors associated with those terms. In these error terms, the subscript $gt$ for a particular parameter $X$ (for instance $X$ is $signal$ or $q_e$) refers to the ground truth data for that parameter $X$ generated in MATLAB using the classical equations and the subscript $L$ for the same parameter $X$ refers to the data generated by the PBML model.

\begin{dmath} 
\label{eqn:eqlabel_LF_3D}
LF = k[(signal_{gt}-signal_{L})^2 + error^2_{voltage}] + l[(q_{e,gt}-q_{e,L})^2 + \Sigma_{z=1}^{Z}(q_{et,gt}-q_{et,L})^2] + n[(q_{h,gt}-q_{h,L})^2 + \Sigma_{r=1}^{R}(q_{ht,gt}-q_{ht,L})^2].
\end{dmath}

The trained weights in the PBML model is compared to that of the ground truth weights, which are the weights used in generating the ground truth data using classical method. We define an error metric as shown in \ref{eqn:errm} which defines how well the trained model weights matches to that of the ground truth weights,
\begin{equation}
\label{eqn:errm}
    Err(w_{eT}) = \sqrt{{\frac{1}{N_{fin}-N_{inj} + 1}} \sum_{i=N_{inj}}^{N_{fin}}\Big\{\frac{w_{eT,lr,i} - w_{eT,gt,i}}{w_{eT,gt,i}} \Big\}^2}.
\end{equation}

\subsection*{Experimental Studies for certain 3D sub-pixels}
\label{3D_results_some_subpixels}
Voxels corresponding to rows and columns ${4,5}$ in \textbf{OX-OY} plane has been used in this simulation experiment. Thus the voxels in 3D volume $(x, y, z)$ corresponding to the subpixels of $2 \times 2$ region in the \textbf{OX-OY} plane are only considered and used in our simulation. Here we consider a virtual boundary outside these voxels and assume that no electrons or holes are injected from outside this virtual boundary at any time $t$. However, it is considered that electrons or holes from these sub-pixels can drift outside this virtual boundary due to the effect of electric field at any time $t$. The electron-hole pairs are injected at position $77$ and $79$ in the \textbf{OZ} direction. Thus, the total of 8 electron-hole pairs are injected at different voxels in this 3D volume. 
\begin{figure*}[htbp]
\begin{multicols}{2}
    \noindent
    \includegraphics[width = \linewidth]{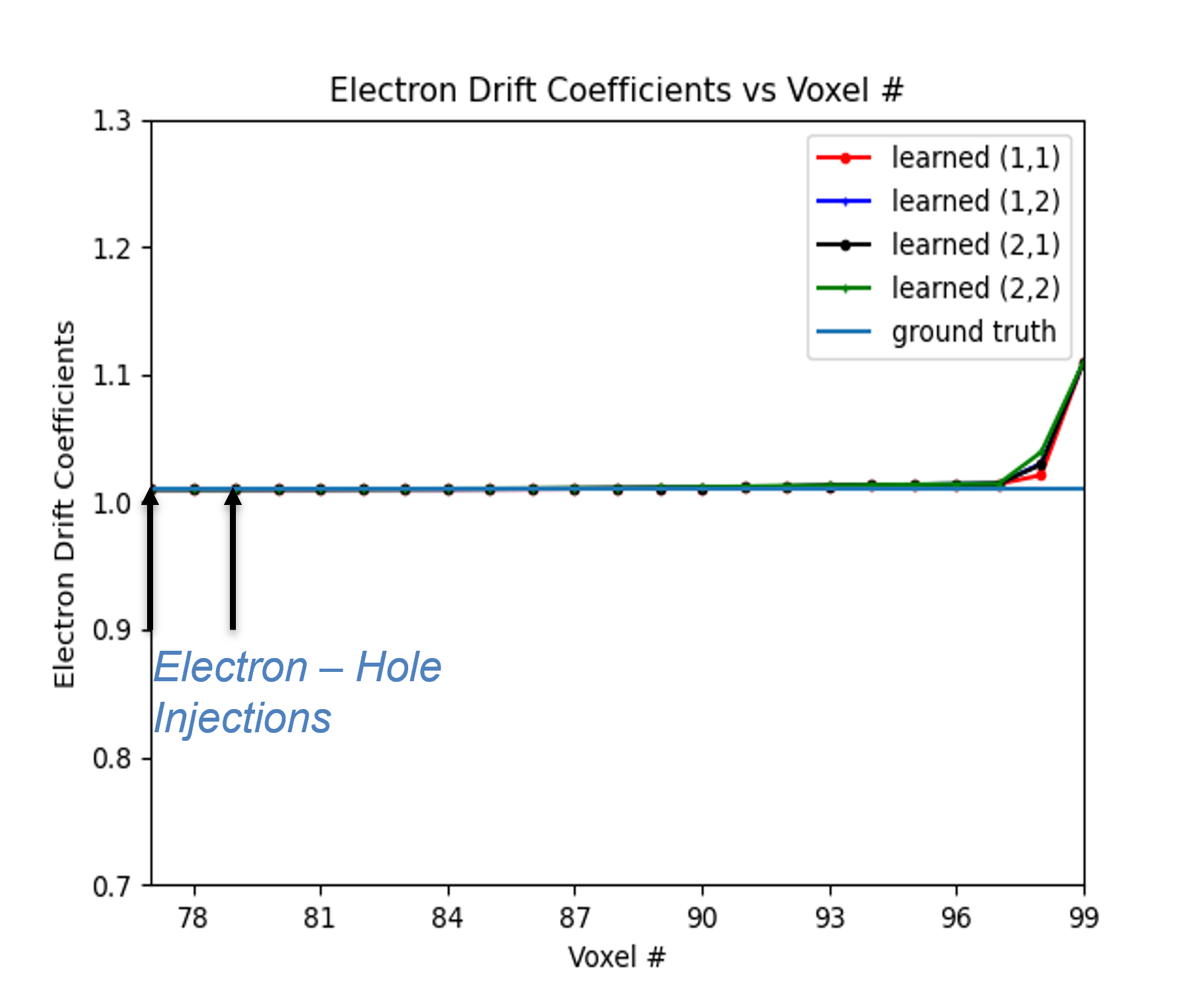}\par\caption*{(a) Electron Drift Coefficients}
    \includegraphics[width = \linewidth]{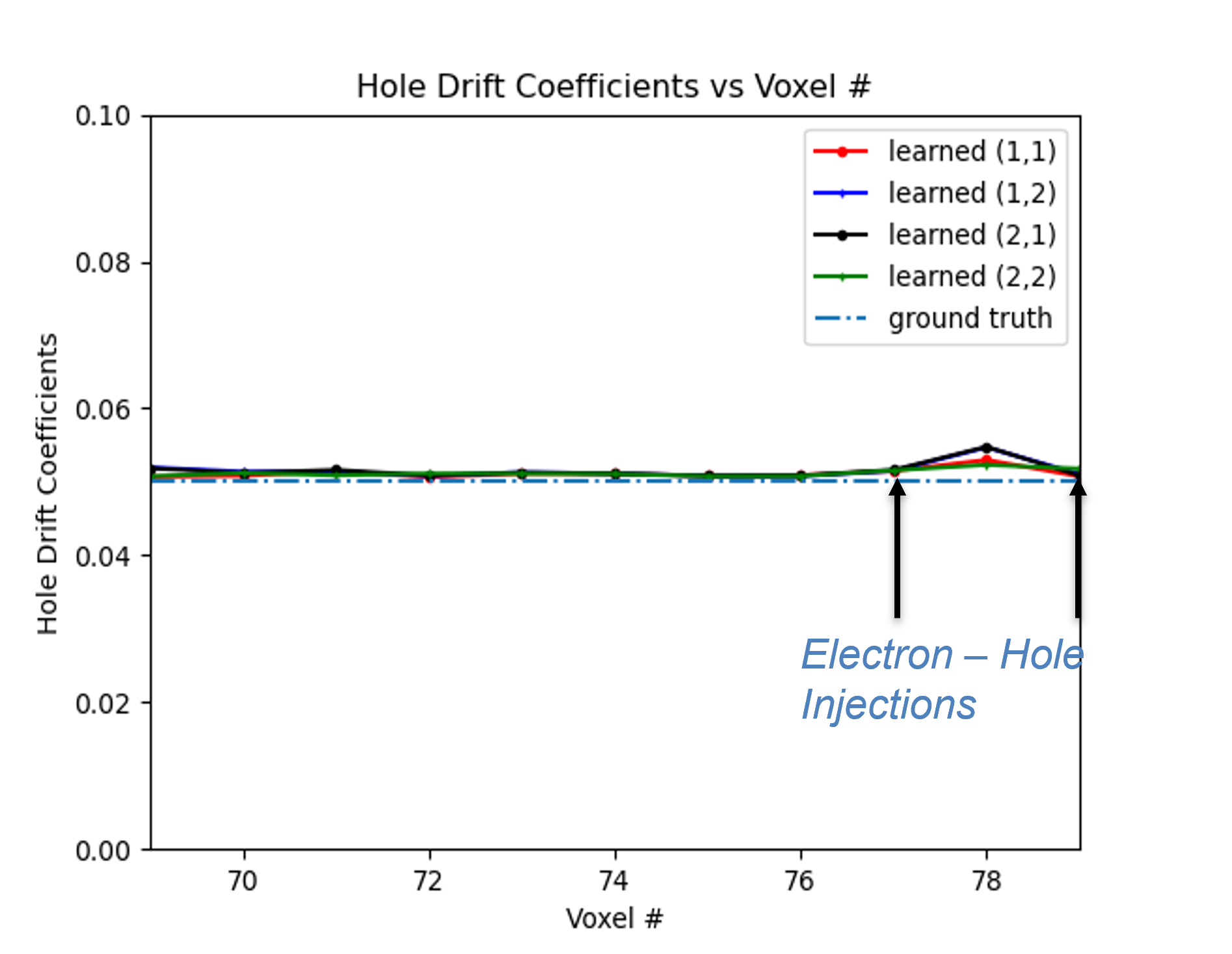}\par\caption*{(b) Hole Drift Coefficients}
\end{multicols}
\caption{Electron and Hole Drift Coefficients for e-h injection at \textbf{OZ} $77$ and $79$. $1$ and $2$ in legend refer to the $4$th and $5$th positions respectively.}
\label{s_subpix_elect_hole_drift}
\end{figure*}

\begin{figure*}[htbp]
\begin{multicols}{2}
    \noindent
    \includegraphics[width = \linewidth]{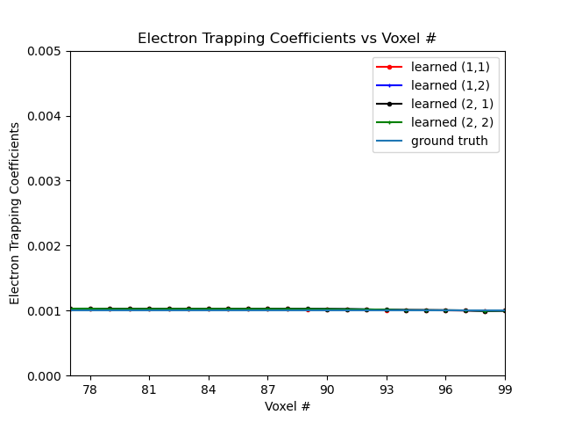}\par\caption*{(a) Electron Trapping Coefficients}
    \includegraphics[width = \linewidth]{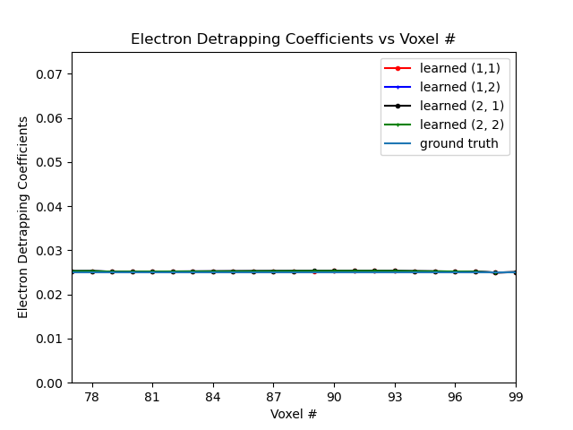}\par\caption*{(b) Electron Detrapping Coefficients}
\end{multicols}
\caption{Electron Trapping and Detrapping Coefficients for e-h injections at \textbf{OZ} positions $77$ and $79$.  $1$ and $2$ in legend refer to the $4$th and $5$th position respectively.}
\label{s_subpix_elect_trap_detrap}
\end{figure*}

\begin{figure*}[htbp]
\begin{multicols}{2}
    \noindent
    \includegraphics[width = \linewidth]{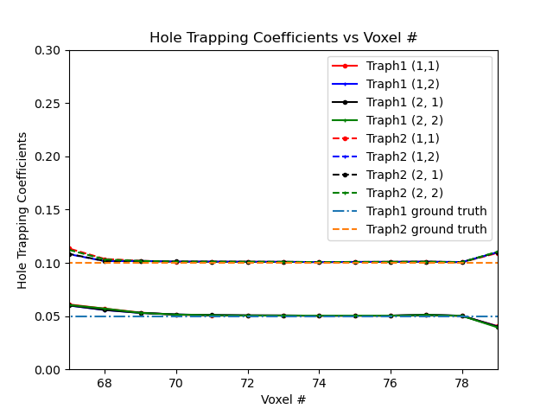}\par\caption*{(a) Hole Trapping Coefficients}
    \includegraphics[width = \linewidth]{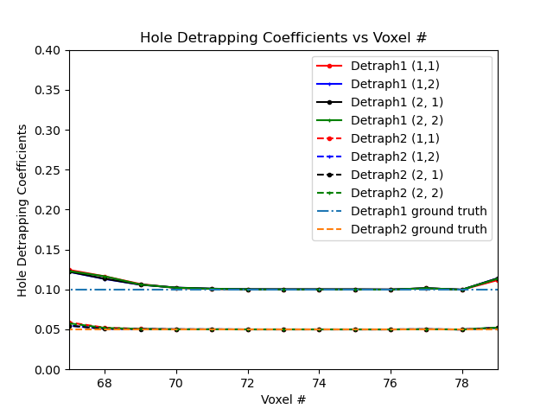}\par\caption*{(b) Hole Detrapping Coefficients}
\end{multicols}
\caption{Hole Trapping and Detrapping Coefficients for e-h injections at \textbf{OZ} $77$ and $79$. $1$ and $2$ in legend refer to the $4$th and $5$th position respectively.}
\label{s_subpix_hole_trap_detrap}
\end{figure*}

Figs. \ref{s_subpix_elect_hole_drift}(a) and (b) shows the variation of electron and hole drift coefficients respectively. The electron drift coefficients are trained from the e-h injection positions all the way to the anode voxel. The hole drift coefficients are trained from the e-h injection positions until \textbf{OZ} positon $69$ which is $9$ positions towards the cathode in \textbf{OZ} direction of the leftmost hole injection \textbf{OZ} position $77$. For the trained PBML model, the $Err(w_{eDrift})$ is $0.0048$ and $Err(w_{hDrift})$ is $0.0322$. Fig. \ref{s_subpix_elect_trap_detrap} (a) and (b) shows the electron trapping and detrapping coefficients. It is observed that the model learns for all the voxel positions perfectly in this subpixels from \textbf{OZ} $77$ to $99$. For electron trapping and detrapping coefficients, $Err(w_{eT})$ is $0.0223$ and $Err(w_{eD})$ is  $0.0098$. For holes, the trapping and detrapping coefficients as shown in Fig. \ref{s_subpix_hole_trap_detrap} learns for $9$ positions from \textbf{OZ} position of $77$, the leftmost hole injection position. For the hole trapping 1 and trapping 2 coefficients for \textbf{OZ} positions of $69$ to $79$, $Err(w_{hT1})$ is $0.0644$ and $Err(w_{hT2})$ is $0.0317$ respectively. On the other hand, for the hole detrapping 1 and detrapping 2 coefficients for \textbf{OZ} positions of $69$ to $79$, $Err(w_{hD1})$ is $0.0452$ and $Err(w_{hD2})$ is $0.0133$.

\begin{figure*}[htbp]
\begin{multicols}{2}
    \noindent
    \includegraphics[width = \linewidth]{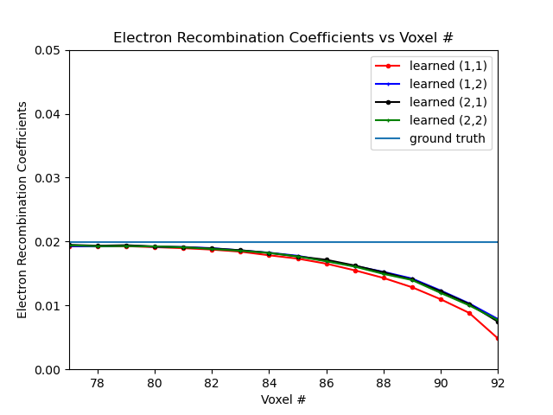}\par\caption*{(a) Electron Recombination Coefficients}
    \includegraphics[width = \linewidth]{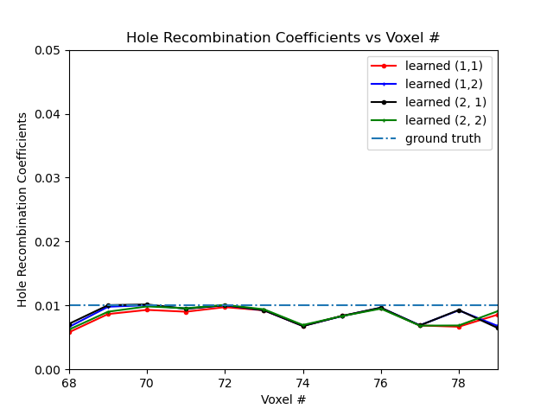}\par\caption*{(b) Hole Recombination Coefficients}
\end{multicols}
\caption{Electron and Hole Recombination Coefficients for e-h injections at \textbf{OZ} positions $77$ and $79$.  $1$ and $2$ in legend refer to the $4$th and $5$th position respectively.}
\label{s_subpix_elect_hole_recomb}
\end{figure*}

Fig. \ref{s_subpix_elect_hole_recomb} (a) and (b) shows the electron and hole recombination coefficients. For the holes, in Fig. \ref{s_subpix_elect_hole_recomb} (b), it is observed that the recombination coefficients are learned close to the ground truth values from \textbf{OZ} positions of $79$ to $69$. For hole recombination coefficients, $Err(w_{hRec})$ is $0.1811$ for \textbf{OZ} positions of $69$ to $79$. However, for the electron recombination coefficients, the coefficients are learned close to the injection positions of \textbf{OZ} of $77$ and $79$. It is observed that beyond position $83$, the error between the learned electron recombination coefficients and ground truth values increases due to convergence of the learned electron recombination coefficients to a different non-optimal value. For the electron recombination coefficients, $Err(w_{eRec})$ is $0.2119$ for \textbf{OZ} positions $77$ to $92$. It is observed that the learned electron recombination coefficients eventually converge to $0$ beyond \textbf{OZ} position of $92$. This behavior of the learned electron recombination coefficients at \textbf{OZ} positions farther away from the e-h injection positions can be attributed to decrease in the gradients away from the e-h injection points. The $Err(w_{eRec})$ calculation thus considers only upto \textbf{OZ} position $92$ from the point of injection. The mean $Err()$ value for the several learned material coefficients in this PBML model is $0.0617$ for the CdZnTe detector based on the e-h injection in \textbf{OZ} of \textbf{OZ} positions $77$ and $79$.

\subsection*{Experimental Studies for all 3D sub-pixels}
\label{3D_results_all_subpixels}
The central pixel volume as shown in Fig. \ref{121_pixels_single_pixel} (right) has been divided into $4 \times 4$ sub-pixels. The simulation data has been generated with anode pixel divided into $12 \times 12$ sub-pixels in \textbf{OX} and \textbf{OY}, as mentioned earlier in this Section. The data (signals, free and trapped charges, electrode weighting potentials) for the $4 \times 4$ sub-pixels has been converted from $12 \times 12$ sub-pixels by simply taking the arithmetic mean of the corresponding data (signals, free and trapped charges, electrode weighting potentials) as $3 \times 3$ sub-pixels in \textbf{OX} and \textbf{OY} in a non-overlapping manner.

\begin{figure*}[htbp]
\begin{multicols}{2}
    \noindent
    \includegraphics[width = \linewidth]{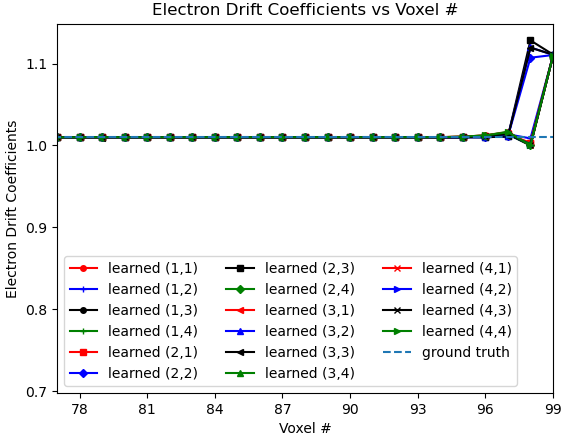}\par\caption*{(a) Electron Drift Coefficients}
    \includegraphics[width = \linewidth]{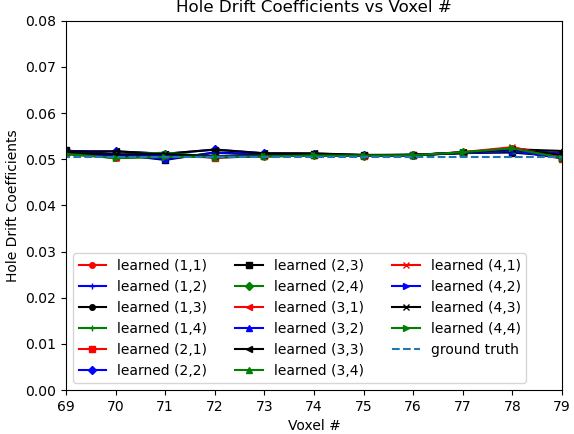}\par\caption*{(b) Hole Drift Coefficients}
\end{multicols}
\caption{Electron and Hole Drift Coefficients in 3D Physical Model for e-h injections at \textbf{OZ} positions $77$ and $79$ for all sub-pixels.}
\label{3D_wTrpt_eh}
\end{figure*}

Fig. \ref{3D_wTrpt_eh} (a) and (b) shows the variation of electron and hole drift coefficients for e-h injection at \textbf{OZ} positions of $77$ and $79$ for all the sub-pixels. The electron drift coefficients are learned from \textbf{OZ} position $77$ to $99$, and the hole drift coefficients are learned from \textbf{OZ} position $69$ to $79$. For the electron drift coefficients, $Err(w_{eDrift})$ is $0.0117$, while for the hole drift coefficients, $Err(w_{hDrift})$ is $0.0152$.  

\begin{figure*}[htbp]
\begin{multicols}{2}
    \noindent
    \includegraphics[width = \linewidth]{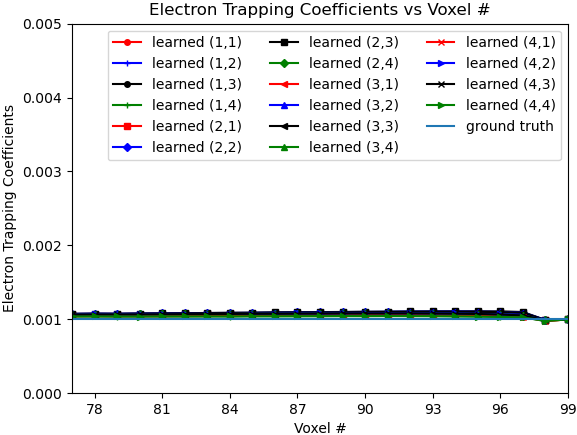}\par\caption*{(a) Electron Trapping Coefficients}
    \includegraphics[width = \linewidth]{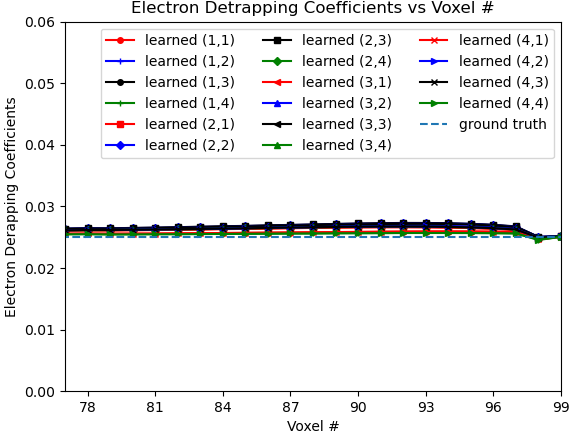}\par\caption*{(b) Electron Detrapping Coefficients}
\end{multicols}
\caption{Electron Trapping and Detrapping Coefficients in 3D Physical Model for e-h injections at \textbf{OZ} positions $77$ and $79$ for all sub-pixels.}
\label{3D_e_wTrap_Detrap}
\end{figure*}

\begin{figure*}[htbp]
\begin{multicols}{2}
    \noindent
    \includegraphics[width = \linewidth]{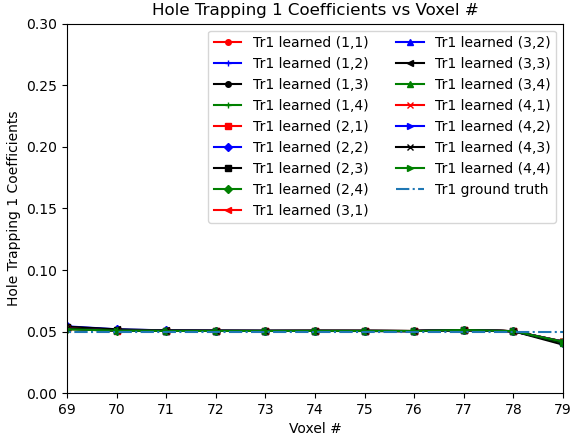}\par\caption*{(a) Hole Trapping 1 Coefficients}
    \includegraphics[width = \linewidth]{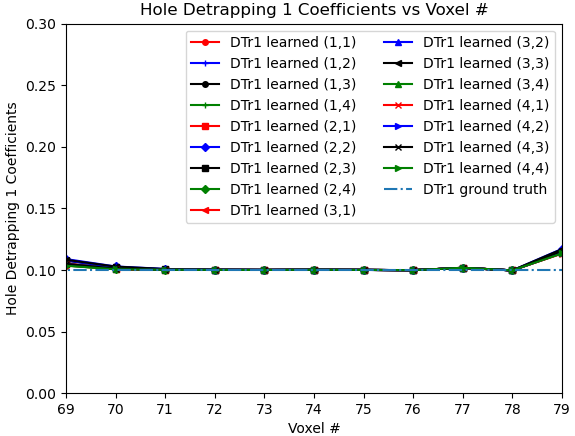}\par\caption*{(b) Hole Detrapping 1 Coefficients}
\end{multicols}
\caption{Hole Trapping 1 and Detrapping 1 Coefficients in 3D Physical Model for e-h injections at \textbf{OZ} positions $77$ and $79$ for all sub-pixels.}
\label{3D_h1_wTrap_Detrap}
\end{figure*}

\begin{figure*}[htbp]
\begin{multicols}{2}
    \noindent
    \includegraphics[width = \linewidth]{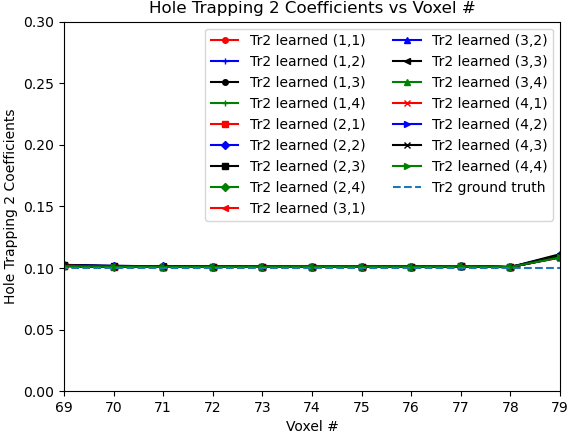}\par\caption*{(a) Hole Trapping 2 Coefficients}
    \includegraphics[width = \linewidth]{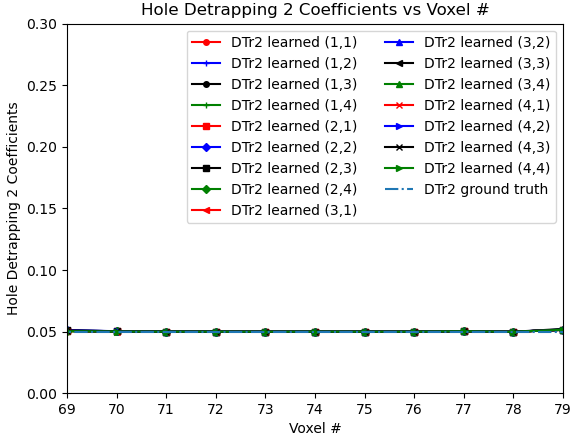}\par\caption*{(b) Hole Detrapping 2 Coefficients}
\end{multicols}
\caption{Hole Trapping 2 and Detrapping 2 Coefficients in 3D Physical Model for e-h injections at \textbf{OZ} positions $77$ and $79$ for all sub-pixels.}
\label{3D_h2_wTrap_Detrap}
\end{figure*}

Fig. \ref{3D_e_wTrap_Detrap} (a) and (b) shows the electron trapping and detrapping coefficients for e-h injection at \textbf{OZ} positions of $77$ and $79$ for all the sub-pixels. The electron coefficients are learned from the leftmost point of e-h injection upto position $99$ in \textbf{OZ} direction. For the electron trapping and detrapping coefficients, $Err(w_{eT})$ is $0.0696$ and $Err(w_{eD})$ is $0.0560$. Similarly, Fig. \ref{3D_h1_wTrap_Detrap}(a) and (b) shows the hole trapping 1 and detrapping 1 coefficients. For hole trapping 1 and detrapping 1 coefficients, $Err(w_{hT1})$ is $0.0228$ and $Err(w_{hD1})$ is $0.0195$ respectively. The hole coefficients are learned from \textbf{OZ} position $69$ to position $79$ for all the sub-pixels. Hole Trapping 2 and Detrapping 2 coefficients are showed in Fig. \ref{3D_h2_wTrap_Detrap} (a) and (b) respectively. For hole trapping 2 and detrapping 2 coefficients, $Err(w_{hT2})$ is $0.0115$ and $Err(w_{hD2})$ is $0.0047$ respectively. For both the hole trapping 1 and 2 alongwith hole detrapping 1 and 2 coefficients, the trained weights converge to the ground truth values. 

\begin{figure*}[htbp]
\begin{multicols}{2}
    \noindent
    \includegraphics[width = \linewidth]{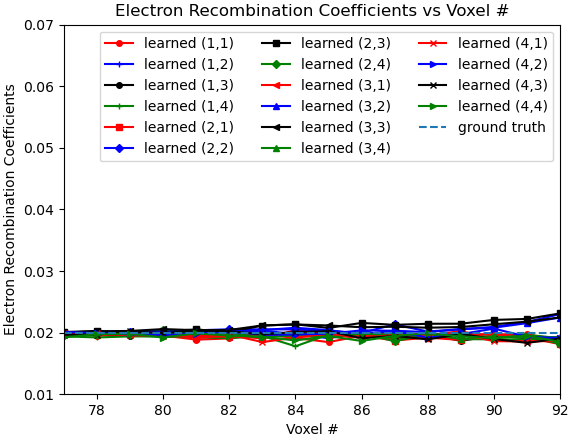}\par\caption*{(a) Electron Recombination Coefficients}
    \includegraphics[width = \linewidth]{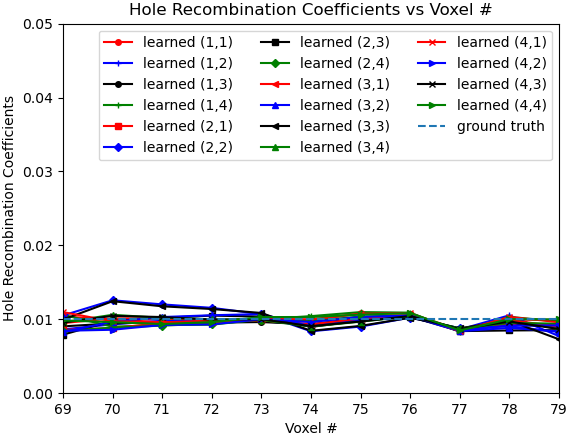}\par\caption*{(b) Hole Recombination Coefficients}
\end{multicols}
\caption{Electron and Hole Recombination Coefficients in 3D Physical Model for e-h injections at \textbf{OZ} positions $77$ and $79$ for all sub-pixels.}
\label{3D_e_h_recomb}
\end{figure*}

Fig. \ref{3D_e_h_recomb}(a) and (b) shows the electron and hole recombination coefficients respectively. For the electron recombination coefficients, the learned coefficients are close to the ground truth at positions close to the injection \textbf{OZ}, while the errors increased as it went further away from the positions of injection towards the anode. For electron recombination coefficients, $Err(w_{eRec})$ is $0.0347$ for \textbf{OZ} position $77$ to $92$. For injection positions $77$ and $79$, the electron recombination coefficients does not converge beyond voxel $92$ due to decrease in gradients in those voxels. For hole recombination coefficents as well, similar effect can be observed, where the learned coefficients from \textbf{OZ} positions of $69$ to $79$, deviate slightly from the ground truth values at \textbf{OZ} positions farther away from the \textbf{OZ} injection positions of $77$ and $79$. For the hole recombination coefficients, $Err(w_{hRec})$ is $0.0873$. For all of these learned material properties using the PBML approach, the mean $Err()$ is $0.0333$.

\section*{Discussion}
\label{3D_Chap_Discussion}
The PBML approach for characterization of the RTSD in 3D is novel and unique. We developed a 3D learning-based model of the RTSD considering the non-uniform electric field and motion of electrons and holes in this non-uniform electric field. CdZnTe has parallel electrode configuration with a single cathode on one end and pixelated anodes on the opposite end. The RTSD volume with a single anode pixel out of $11 \times 11$ pixels and a single cathode on the opposite end of the anodes has been considered in this simulation study. The single pixel has been sub-pixelized in 2D, \textbf{OX} and \textbf{OY} directions. Further subdivision has been done in \textbf{OZ} direction. The resulting sub-divided volumes of the crystal are termed as voxels. In each voxel, the non-uniform motion of charge particles, alongwith trapping, detrapping and recombination of charges are modeled as learnable parameters. The model has been trained in a supervised manner using the input-output ground truth data generated using classical equations. The physics based learning model gets trained by using a back-propagation algorithm which updates the trainable model weights in each epochs. 

Experimental simulations has been done considering a subset of sub-pixels in the 3D model and also considering all the sub-pixels in the 3D model. For electron-hole pair injections at different \textbf{OZ} positions, the charge transport, trapping and detrapping coefficients converge to the ground truth values with minimum error. The hole recombination coefficients also converge to the ground truth values for the positions where the holes drift. However, from our experimental studies we observe that the electron recombination coefficients gets trained for only $14$ positions from the left-most electron-hole injection position. Moreover, for the experiments with a set of sub-pixels, the convergence of the electron recombination coefficients are poorer than that for the experiments considering all the sub-pixels. The $Err()$ value for the experiment done with a subset of sub-pixels is higher than that of the case with all sub-pixels, which is attributed primarily due to non-convergence of the electron recombination coefficients to the ground truth parameters for the former experiment. The lack of convergence of the electron recombination coefficients for positions far away from the position of injection will be investigated in the future work.  

The robustness of the 3D learning-based model has been tested by using 3D uniform electric field and electron/hole charge motion in that field. Despite the fact that the model has incorporated non-uniform charge motion, the model correctly learns the uniformity of the charge motion and identifies that there is no second order drift of charges. The physics based learning model also identifies the correct material properties - transport, trapping, detrapping and recombination at a finer scale. Experiments with charge motion in 3D non-uniform electric field will be performed as a future work as well.

\begin{figure}
  \centering
  \includegraphics[width=0.8\linewidth]{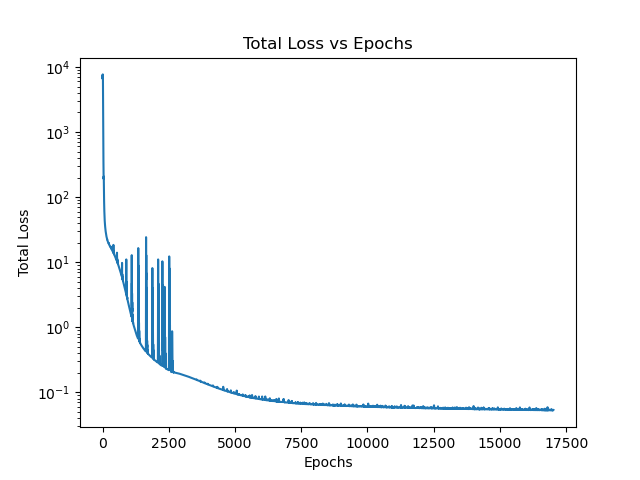}
  \caption{Variation of loss over epochs.}
\label{s_subpix_Loss_epochs}
\end{figure}

\section*{Methods}

In order to train and test the PBML model, actual measured data would be needed. Unfortunately, no such dataset is available in the literature for RTSD. In order to generate this dataset, we generate synthetic data using the classical equations in MATLAB using pre-defined material properties which are the ground truth parameters. The dataset consists of electron-hole pairs injected at different voxels of the 3D model and the corresponding signals at the electrodes over time along with the electron and hole charges (free and trapped) in the voxels over time. The magnitude of the injected charges are normalized to $1$. Following the charge conservation in the bulk of the RTSD, the magnitude of the free and trapped charges are always less than $1$. The signals at the electrodes have the range of $[-1, 1]$. This range of signals and charges (free and trapped) are due to the combination of material properties chosen apriori. Since the learning-based model is developed in a voxelized manner, the training data generated using the classical approach is also voxelized. The experimental data using the classical model has been developed for non-uniform electric weighting potential between the cathode and anode pixels, and also based on the location of the anode pixels. During training, the PBML model is trained using only a subset of data from the dataset as the charges drift to neighboring voxels from the injection positions and the weights in those voxels are trained in this process.

The model weights are initialized during the start of training process. The model is trained over several epochs by computing the loss function based on the output corresponding to each input injections for the different reduced models. The model is a recurrent network structure over time, and hence Backpropagation through Time (BPTT) \cite{rumelhart1985learning,werbos1990backpropagation} is used to compute the gradients of the loss with respect to the trainable weights in the model. Stochastic gradient descent based method - ADAM \cite{kingma2014adam} is used for optimization and the weights are updated in each epoch. The learning rate is initialized at $5 \times 10^{-4}$ with $2$ momentum terms set as $\beta_{1}=0.9$ and $\beta_{2}=0.999$. The training loss reduces over epochs. However, after certain epoch $E_{1}$, the training loss started oscillating. These oscillations increases with epochs. After observing this oscillation for certain epochs $E_{2}$, the learning rate was reduced to $1 \times 10^{-5}$ which resulted in better convergence of the loss function. 

\begin{figure}
  \centering
  \includegraphics[width=0.8\linewidth]{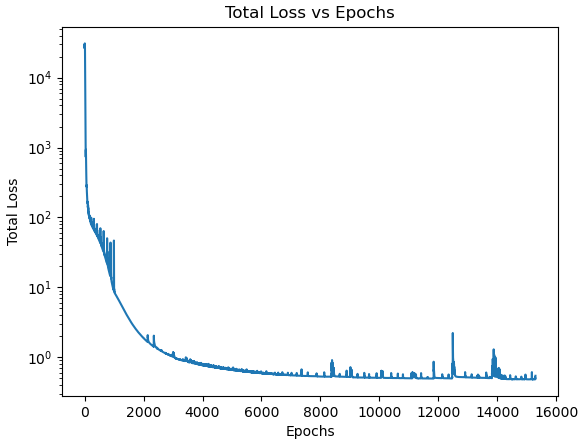}
  \caption{Variation of loss over epochs.}
\label{3D_model_Loss_epochs}
\end{figure}

For the experimental study with certain 3D sub-pixels, the variation of loss over epochs is shown in Fig. \ref{s_subpix_Loss_epochs}. It is seen that the training loss starts to oscillate from epoch $500$ which keeps on increasing. However, when the learning rate is reduced at epoch $2500$, it is seen that the training loss stops oscillating and training loss further reduces until it saturates and stops reducing any further. Fig. \ref{3D_model_Loss_epochs} shows the variation of training loss over epochs for the experimental studies with all 3D sub-pixels in a single pixel. At epoch $1000$ the learning rate is changed to $1 \times 10^{-5}$ which results in no significant oscillation of the training loss.

Once the model is trained, the weights of the model converges to the detector ground truth parameters. Our model has been developed using the popular machine learning Tensorflow library \cite{tensorflow2015-whitepaper} in Python using eager execution mode. During training, the loss is monitored over epochs and allowed to converge until it stops reducing. In our experiments, there is no improvement in the trained weights of this PBML model when there is no reduction in training loss. We performed experimental studies with $k=1$ and $l=n=1000$ in Eqn. \ref{eqn:eqlabel_LF_3D} for a model having $100$ voxels. The single anode pixel (out of $11 \times 11$ pixels) was divided into $12 \times 12$ sub-pixels, with the cathode as a single electrode and then the data for signals, free and trapped charges has been generated with electron-hole injection at each positions.

\section*{Data Availability}
All the data supporting the finds of this work will be shared upon reasonable request to the authors. Correspondence and requests for data should be addressed to S.B..

\section*{Code Availability}
The code for the PBML model supporting the finds of this work will be shared upon reasonable request to the authors.

\section*{Author Contributions}
S.B., M.R. and A.K.K. conceived the idea. S.B. implemented the idea and conducted all the simulations. All authors analyzed the results. S.B. and M.B. wrote the manuscript. All authors reviewed and approved the final version of the manuscript.

\bibliography{main}

\section*{Acknowledgment}
The authors acknowledge Siemens Medical Solutions, Inc. for providing financial support to Northwestern University for conducting this research within scope of a research agreement.

\section*{Competing Interests}
The authors declare no competing interests.

\end{document}